\documentclass{iau} 
\usepackage{graphicx}
\usepackage{hyperref}

\title[Dynamo Processes Constrained by Observations] 
{Dynamo Processes Constrained by Solar and Stellar Observations}

\author[Maria A. Weber]   
{Maria A. Weber$^{1,2}$
}

\affiliation{$^1$Department of Astronomy and Astrophysics, University of Chicago 
 \\[\affilskip] 
$^2$Department of Astronomy, Adler Planetarium, Chicago, IL \\email: {\tt maweber@uchicago.edu}}

\pubyear{2018}
\volume{340}  
\jname{Long Term Datasets for the Understanding of Solar and Stellar Magnetic Cycles}
\editors{D. Banerjee, J. Jiang, K. Kusano, \& S. Solanki, eds.}
\begin{document}

\maketitle

\begin{abstract}
Our understanding of stellar dynamos has largely been driven by the phenomena we have observed of our own Sun. Yet, as we amass longer-term datasets for an increasing number of stars, it is clear that there is a wide variety of stellar behavior. Here we briefly review observed trends that place key constraints on the fundamental dynamo operation of solar-type stars to fully convective M dwarfs, including: starspot and sunspot patterns, various magnetism-rotation correlations, and mean field flows such as differential rotation and meridional circulation. We also comment on the current insight that simulations of dynamo action and flux emergence lend to our working knowledge of stellar dynamo theory. While the growing landscape of both observations and simulations of stellar magnetic activity work in tandem to decipher dynamo action, there are still many puzzles that we have yet to fully understand. 

\end{abstract}

\firstsection 
\section{Introduction: Our Unique Dynamo Perspective}
Historically, our study of stellar dynamos has been shaped by observations of the Sun over a small portion of its existence. Magnetic braking facilitated through the solar wind has spun down the Sun from a likely chaotic youth to its current middle-aged state. Now it exhibits an $\sim$eleven year cycle of sunspot activity, with bands of flux emergence that migrate equatorward from mid-latitudes.  The global-scale structure of the Sun's coronal field also changes during this cycle. Furthermore, helioseismology has shown that the convection zone does not rotate uniformly. \textit{Ultimately, it is the nature of the turbulent, rotating convection inside the Sun that generates and sustains its magnetic properties.}          

How might our knowledge of stellar dynamos have evolved if we lived around a different star? For example, the K dwarf HAT-P-11 has a similar rotation and activity cycle period as the Sun, including active latitudes within $\pm$45$^{\circ}$ (\cite[Morris \etal\  2017]{Morris_etal17a}). With its low-lying starspot bands and outer convection zone, we still might have hypothesized that toroidal bands of magnetism were generated in the tachocline at the radiative interior/convection zone interface - the long-held solar dynamo paradigm. But, what if we orbited a fully convective M dwarf ($\lesssim$0.35M$_{\odot}$)? Without tachoclines, late M dwarfs are among the most magnetically active, with frequent flares that rival those of the Sun (\cite[e.g. Yang \etal\ 2017]{Yang_etal17}). Many are also highly spotted at all latitudes, including near the poles (\cite[e.g. Barnes \etal\ 2015]{Barnes_etal15}).  The constraints placed on dynamo theory by long-term observations of such host stars might have led us to a different way of solving the stellar dynamo problem.


Motivated by the varieties of stellar magnetism, we present a brief review of the ways by which the most recent observations of main-sequence F to M-type stars constrain dynamo theory. We also highlight insight that current magnetohydrodynamic (MHD) simulations provide in deciphering observed solar and stellar dynamo behavior. In what follows, we focus on mean flows such as differential rotation and meridional circulation (Sec. \ref{sec:meanflows}), the many forms of magnetism-rotation correlations (Sec. \ref{sec:magrot}), and sunspot and starspot trends (Sec. \ref{sec:spottrends}). Across these categories are further questions as to what establishes the strength of the magnetism built, the role of the tachocline, and the generation and rise of starspot progenitors. For a more in-depth discussion of the solar-stellar connection, we direct the reader toward the review by \cite[Brun $\&$ Browning 2017]{BrunBrowning_17}.   

\section{Mean Flows: Differential Rotation and Meridional Circulation}
\label{sec:meanflows}
Stellar magnetism is generated and sustained by plasma fluid motions. Mean flows such as differential rotation and meridional circulation both play integral roles in transporting magnetic flux and redistributing angular momentum within the convection zone.

{\underline{\bf Differential Rotation}}: Helioseismology has revealed a conical solar angular velocity profile with a prograde equator and slower rotating poles. There are two layers of strong shear; the tachocline at the radiative interior/convection zone interface and the near-surface shear layer across the upper $\sim0.05$R$_{\odot}$ (\cite[e.g. Howe 2009]{Howe_09}). The role both shearing regions play in shaping solar magnetism is still unclear. For other solar-like stars, surface differential rotation can be monitored by tracking photometric starspot signatures. Such observations show that differential rotation increases weakly with stellar effective temperature, then strongly near the transition to F-type stars (\cite[Reinhold $\&$ Gizon 2015]{reindhold_15}). 

Global MHD simulations have made contact with these observed trends. Differential rotation contrasts are found to increase in F-type stars compared to M-dwarfs, as well as with rapid rotation (\cite[see Brun $\&$ Browning 2017]{BrunBrowning_17}). Most tend toward cylindrical angular velocity contours, following the so-called Taylor-Proudman constraint (\cite[e.g. Featherstone $\&$ Miesch 2015]{Featherstone_etal15}). A more solar-like, conical profile can be achieved by upsetting the balance of forces that regulate angular momentum transport. This can be done by imposing a latitudinal entropy gradient to mimic coupling across the tachocline interface through a thermal wind balance (\cite[Miesch \etal\ 2006]{miesch_etal06}), or by sufficient Reynolds stresses and Lorentz forces introduced by the presence of magnetism (\cite[e.g. Guerrero \etal\ 2016]{guerrero_etal16}). In simulations a clear transition to anti-solar differential rotation (slow equator, fast poles) occurs at slower rotations, near a convective Rossby number of 1 (\cite[Gastine \etal\ 2014]{Gastine_etal14}). The Rossby number captures the relative importance of convection verses rotation. The presence of dynamo action may alter these regime transitions (\cite[e.g. Karak \etal\ 2015]{Karak_etal15}).            

{\underline{\bf Meridional Circulation}}: There is still a debate as to the structure of the solar meridional circulation pattern. Some helioseismology studies suggest that it may be multi-celled in radius (\cite[Zhao \etal\ 2013]{Zhao_etal13}), rather than a single cell in both hemispheres that is poleward at the surface and equatorward in the deep interior. In global-scale hydrodynamic simulations, a multi-celled meridional circulation pattern appears as the star is spun up (\cite[e.g.  Featherstone $\&$ Miesch 2015]{Featherstone_etal15}). This behavior is linked to a change in the convective Reynolds stresses as rotational constraints become more pronounced, and is indicative of the phenomenon of gyroscopic pumping. The meridional circulation kinetic energy (and therefore the speed of the flow) also decreases with increasing rotation (\cite[see Brun $\&$ Browning 2017]{BrunBrowning_17}), and corresponds with the increase in differential rotation noted above. 

\section{Myriad Magnetism-Rotation Correlations}
\label{sec:magrot}
There are many flavors of magnetism-rotation trends that reveal themselves in observations, placing key constraints on dynamo theory. Below we highlight a few examples, including the `canonical' activity-rotation correlation, hints as to what sets the strength of the magnetism built, the cycle period and variability, and global-scale field topology.      

{\underline{\bf Activity-Rotation Trends}}: Rapidly rotating stars tend to have a greater level of magnetic activity. This phenomenon appears when plotting the X-ray to bolometric luminosity ratio (a proxy for magnetic activity) as a function of Rossby number. There is a strikingly similar activity-rotation correlation among F to M-type stars; activity increases with rapid rotation, and then saturates upon reaching a Rossby number threshold (\cite[e.g. Wright \etal\ 2013]{Wright_etal13}). This correlation is the same for both partially and fully convective M dwarfs (\cite[Wright $\&$ Drake 2016]{Wright_etal16}). In these formulations, the Rossby number is given as the ratio of rotation period to the convective turnover timescale taken from mixing length models ($Ro=P_{rot}/\tau$). Recently, there has been a push to move away from such treatments requiring $\tau$ in favor of better constrained quantities (\cite[e.g. Reiners \etal\ 2014]{Reiners_etal14}). 

Interestingly, the fraction of active M dwarfs actually increases across the `tachocline divide' toward fully convective stars (\cite[e.g. Schmidt \etal\ 2015]{Schmidt_etal15}). Of the smaller fraction of active early M dwarfs, most tend to rotate slower than their active later-type counterparts (\cite[Jeffers \etal\ 2018]{Jeffers_etal18}). This signals a change in the way the stars lose angular momentum, which is linked to their age, mass-loss rate, and large-scale magnetic field topology.       

{\underline{\bf Strength of Magnetism Built}}: Dynamo simulations give us some indication as to what sets the magnetism built inside stars, and therefore the strength of the activity. One simple estimation is that the mean magnetism is in equipartition with the kinetic energy density of convection. Another assumption is that the thermal energy flux that convection carries can be converted to magnetic energy. A scaling law based on this approach by \cite[Christensen \etal\ (2009)]{Christensen_etal09} matches well with the magnetism generated in geodynamo simulations and observations of some stars and planets, but it overestimates the magnetism observed in slower rotating solar-like stars. As the landscape of convective dynamo simulations grows, it can now be shown in some regimes that the ratio of magnetic energy to kinetic energy increases with rapid rotation (\cite[Viviani \etal\ 2017]{Viviani_etal17}). Through a combination of Ohmic dissipation and magnetic buoyancy arguments (in the context of fully convective M dwarfs), \cite[Browning \etal\ (2016)]{Browning_etal16} show that there may be an upper limit to the magnetism that can be achieved by dynamo action.     



{\underline{\bf Cycle Period and Variations}}: Many observational studies have attempted to classify a relationship between stellar cycles and rotation periods. \cite[B\"ohm-Vitense (2007)]{Bohm_07} (among others) find two `branches' of activity in cycle period/rotation period space wherein activity increases with rotation, with the Sun nestled in the space between the so-called active and inactive branches. \cite[Metcalfe $\&$ van Saders (2017)]{Metcalfe_etal17} suggest that the position of the Sun in such plots is linked to its evolutionary track, and is a natural consequence of magnetic braking as it ages. There is also some indication that some young, fast rotating solar analogs have an amplitude of variability many times that of the solar cycle, while slow rotators have little variability (\cite[e.g. Egeland \etal\ 2017]{Egeland_etal17}). 

Progress is being made to determine the processes that set the cycle period as a function of stellar parameters. Using a class of simulations called Babcock-Leighton Flux Transport models, it is found that the cycle period actually lengthens at faster rotations (\cite[Jouve \etal\ 2010]{Jouve_etal10}). Broadly, such models solve the magnetic induction equation while prescribing the fluid flows that advect the magnetism, where the meridional circulation often plays a primary role. In \cite[Jouve \etal\ (2010)]{Jouve_etal10}, the cycle period relationship results from a decrease in the speed of the meridional circulation as the rotation increases (see Sec. \ref{sec:meanflows}). Some global models of convective dynamo action also find that the cycle period is inversely proportional to the rotation rate (\cite[Strugarek \etal\ 2017]{Strugarek_etal17}). These results are compatible with observations when the stellar luminosity (linked to the convective vigor) is taken into account. Similar simulations exhibit grand minima states, akin to the solar Maunder minimum (\cite[Augustson \etal\ 2015]{Augustson_etal15}). This arises in part from the low magnetic Prandtl number achieved (the ratio of viscous to magnetic diffusivities), resulting in a highly time-dependent behavior. Incorporating a tachocline in such models might extend the cycle length, with the timescale regulated by the diffusion of the toroidal magnetic field generated in the stable layer (\cite[Guerrero \etal\ 2016]{Guerrero_etal16}).         

{\underline{\bf Global-Scale Field Topology}}:
The global-scale field topology also changes with various stellar parameters. Using the technique of Zeeman-Doppler imaging, \cite[See \etal\ 2016]{See_etal16} find that stars with $Ro\gtrsim1$ tend to have mostly poloidal and axisymmetric fields, whereas $Ro\lesssim1$ stars can generate toroidal fields with non-axisymmetry. By analyzing the link between Stokes-I (unpolarized light) and Stokes-V (circularly polarized) spectral lines of a selection of M dwarfs, \cite[Shulyak \etal\ 2017]{Shulyak_etal17} recently discovered that the stars that generated the strongest fields ($\gtrsim4$ kG) tend to be fully convective, rapid rotators with mostly dipolar fields. An important note is that a bistability exists where M dwarfs of similar masses and rotation can show either dipolar or multipolar geometries.

Global dynamo simulations provide some clues as to what sets the large-scale topology of stars. It is common to find toroidal wreaths of magnetism built within the convection zone of rapidly rotating, solar-like stars (e.g. \cite[Nelson \etal\ 2014]{Nelson_etal14}, \cite[Augustson \etal\ 2015]{Augustson_etal15}). Within M dwarf models, there is a bistability of both dipolar and multipolar dominated dynamos at low local Rossby number (\cite[e.g. Gastine \etal\ 2013]{Gastine_etal13}), perhaps shedding light on the observations above. This behavior can be understood in a few ways; one is driven by the way magnetism adjusts the angular momentum balance. The dipole-dominated solutions tend to have weak differential rotation, and could be classified as $\alpha^{2}$ dynamos where small-scale turbulence generates both toroidal and here much stronger poloidal fields. Multipolar-dominated solutions tend to have significant differential rotation, with the production of toroidal fields by this shear (the $\Omega$-effect) playing a larger role.    


\section{Sunspot and Starspot Trends}
\label{sec:spottrends}
Starspots are windows into a star's deep seated dynamo mechanism, encoding information about the generation of this magnetism and its rise to the surface. In particular, solar dynamo simulations are \textit{heavily} constrained by sunspots.

{\underline{\bf How do stars get their spots?}}: Often the flux tube model is invoked to describe the evolution of magnetic field bundles presumed to be the progenitors of starspots (\cite[see Fan 2009]{Fan_09}). These models assume that either shear within the convection zone or in the tachocline can generate fibril, toroidal magnetism that rises under the effects of rotation, convection, and buoyancy. Properties are extracted once the simulated buoyant loop reaches the near surface, and compared statistically to observations (\cite[e.g. Weber \etal\ 2013b]{Weber_etal13b} in the solar context). Only recently have convective dynamo simulations been able to self-consistently capture elements of magnetic flux emergence (\cite[Nelson \etal\ 2014]{Nelson_etal14}, \cite[Fan $\&$ Fang 2014]{Fan_etal14}). These rising magnetic structures are built in the lower-to-mid convection zone in dynamo models that achieve a particularly low level of diffusion. In the context of fully convective stars, strong dipolar fields can locally suppress convection, creating a polar starspot (\cite[Yadav \etal\ 2015]{Yadav_etal15}). Recent flux tube simulations in M dwarfs can also account for simultaneous high and low latitude starspots (\cite[Weber $\&$ Browning 2016]{Weber_etal16}). 


{\underline{\bf Tilt Angles, Longitudes, and Latitudes}}: Solar MHD simulations need to explain observed sunspot trends, including: equatorward moving sunspot bands confined to $\pm35^{\circ}$ latitudes, the tilting of sunspot pairs toward the equator (i.e. tilt angles) which increases with emergence latitude (Joy's Law), and longitudinal bands/nests of emergence often referred to as active longitudes. There is a growing landscape of starspot observations, although with far less fidelity than we can achieve for the Sun. Spots have been observed at all latitudes, with some stars also displaying active longitudes (\cite[see Strassmeier 2009]{Strassmeier_09}). From flux emergence simulations, we have learned that sunspot tilt angles arise from a combination of the magnetic field line twist, the Coriolis force, and helical convective upflows acting on rising bundles of magnetic field (\cite[e.g. Fan 2009]{Fan_09}, \cite[Weber \etal\ 2013b]{Weber_etal13b}). Active longitudes might result from instabilities of the tachocline and the magnetism built therein (\cite[Dikpati $\&$ Gilman 2005]{Dikpati_etal05}), the localization of super-equipartition fields established by the dynamo (\cite[e.g. Nelson \etal\ 2014]{Nelson_etal14}), or the presence of giant cell convection which forms `windows' within which flux prefers to emerge (\cite[Weber \etal\ 2013a]{Weber_etal13a}). High latitude spots observed on some rapidly rotating stars might arise from deflection of the flux tubes toward the poles due to the Coriolis force (\cite[Sch\"ussler \etal\ 1996]{Schuessler_etal96}), or again because of the presence of strong dipolar fields (\cite[Yadav \etal\ 2015]{Yadav_etal15}).


{\underline{\bf Relation to Flux Transport Models}}: For the Sun, there is strong observational evidence that the emergence and decay of active regions is related to its polarity reversal (\cite[e.g. Babcock 1961]{Babcock_61}). In Babcock-Leighton Flux Transport models, active region tilt angles serve as the source for the poloidal field. By incorporating a scatter around the mean tilt angle trend, \cite[Karak \& Miesch (2017)]{Karak_etal17} illustrate that solar cycle variability can be attributed to tilt angle fluctuations alone, resulting in periods of grand minima and maxima. The question remains as to whether a Babcock-Leighton Flux Transport model might be in operation on other stars, especially rapid rotators that exhibit high-latitude starspots, fully convective stars that are highly spotted, or the Sun at a different age. 

{\underline{\bf Equatorward Propagation}}: The propagation of magnetic bands in spherical, global dynamo simulations is usually toward the poles, opposite the solar trend. Typically these models obey the Parker-Yoshimura rule, which relates the direction of dynamo wave propagation to the sign of the kinetic helicity of the flow and the radial differential rotation gradient (\cite[see e.g. Brun $\&$ Browning 2017]{BrunBrowning_17}). Although the Parker-Yoshimura rule is satisfied, \cite[K\"apyl\"a \etal\ (2013)]{Kapyla_etal13} find a set of simulations that transition to equatorward motion as the density stratification is increased. The effect is to shift the location of the processes that generate magnetism by shearing ($\Omega$-effect) and small-scale turbulence ($\alpha$-effect). However, simulations by \cite[Augustson \etal\ (2015)]{Augustson_etal15} show equatorward propagation that does not follow the Parker-Yoshimura rule.  In that case, there is a tight correlation between the presence of toroidal magnetism and shear. Strong shearing regions move equatorward as Lorentz forces of magnetic wreaths locally weaken the shear over time.    



\section{Summary and Outlook}



Simulations of dynamo action and flux emergence are approaching contact with many aspects of observed stellar magnetism and convection. Although, we emphasize that the models discussed here are far from matching the parameters of actual stellar interiors. Broadly, we know that convection, rotation, and their relative strengths set many trends in stellar magnetism. The nature of the established mean convective flows has an impact on the resulting dynamo action over various spatial and temporal scales. Rotation plays a role in setting stellar activity levels, the cycle period, and the global-scale field topology. Some aspects of sunspot and starspot properties have been explored and replicated with models, but only now are dynamo simulations showing some hints of self-consistent flux emergence. There is a long list of other dynamo-constraining observational trends that we have neglected, opting to focus on those of closer alignment with the scope of IAUS 340. As we survey the landscape of stellar dynamo theory, we note that there are still many mysteries yet to be pieced together, and look forward to future observations that further constrain (and complicate) our understanding of stellar magnetism.  

{\underline{\it Acknowledgements}}: M.W. is supported by a National Science Foundation Astronomy and Astrophysics Postdoctoral Fellowship (AST-1701265). She thanks the IAUS 340 organizers for inviting her to give a presentation with the same title as the review here.


\begin{thebibliography}{}

\bibitem[Augustson \etal\ (2015)]{Augustson_etal15}
{Augustson, K., Brun, A.S., Miesch, M., \& Toomre, J.} 2015,
\href{http://adsabs.harvard.edu/abs/2015ApJ...809..149A}{\textit{ApJ}, 809, 149}

\bibitem[Babcock \etal\ (1961)]{Babcock_61}
{Babcock, H.W.} 1961,
\href{http://adsabs.harvard.edu/abs/1961ApJ...133..572B}{\textit{ApJ}, 133, 572} 

\bibitem[Barnes \etal\ (2015)]{Barnes_etal15}
{Barnes, J.R., Jeffers, S.V., Jones, H.R.A., Pavlenko, Ya. V., Jenkins, J.S., Haswell, C.A., \& Lohr, M.E.} 2015,
\href{http://adsabs.harvard.edu/abs/2015ApJ...812...42B}{\textit{ApJ}, 812, 42}

\bibitem[Bohm-Vitense (2007)]{Bohm_07}
{B\"ohm-Vitense, E.} 2007,
\href{http://adsabs.harvard.edu/abs/2007ApJ...657..486B}{\textit{ApJ}, 657, 486}

\bibitem[Brun \etal\ (2017)]{BrunBrowning_17}
{Brun, A.S., $\&$ Browning, M.K.} 2017,
\href{http://adsabs.harvard.edu/abs/2017LRSP...14....4B}{\textit{LRSP}, 14, 4}

\bibitem[Browing \etal\ (2016)]{Browning_etal16}
{Browning, M.K., Weber, M.A., Chabrier, G., \& Massey, A.P.} 2016,
\href{http://adsabs.harvard.edu/abs/2016ApJ...818..189B}{\textit{ApJ}, 818, 189} 


\bibitem[Christensen \etal\ (2009)]{Christensen_etal09}
{Christensen, U.R., Holzwarth, V., \& Reiners, A.} 2009,
\href{http://adsabs.harvard.edu/abs/2009Natur.457..167C}{\textit{Nature}, 457, 167} 



\bibitem[Dikpati \etal\ (2005)]{Dikpati_etal05}
{Dikpati, M., \& Gilman, P.A.} 2005,
\href{http://adsabs.harvard.edu/abs/2005ApJ...635L.193D}{\textit{ApJL}, 635, L193} 

\bibitem[Egeland \etal\ (2017)]{Egeland_etal17}
{Egeland, R., Soon, W., Baliunas, S., Hall, J.C., \& Henry, G.W.} 2017,
\href{http://adsabs.harvard.edu/abs/2017IAUS..328..329E}{\textit{Proc. IAUS 328}, 328, 329}

\bibitem[Fan (2009)]{Fan_09}
{Fan, Y.} 2009,
\href{http://adsabs.harvard.edu/abs/2009LRSP....6....4F}{\textit{LRSP}, 6, 4}

\bibitem[Fan \etal\ (2014)]{Fan_etal14}
{Fan, Y., \& Fang, F.} 2014,
\href{http://adsabs.harvard.edu/abs/2014ApJ...789...35F}{\textit{ApJ}, 789, 35}

\bibitem[Featherstone \etal\ (2015)]{Featherstone_etal15}
{Featherstone, N.A., \& Miesch, M.S.} 2015,
\href{http://adsabs.harvard.edu/abs/2015ApJ...804...67F}{\textit{ApJ}, 804, 67} 

\bibitem[Gastine \etal\ (2013)]{Gastine_etal13}
{Gastine, T,. Morin, J., Duarte, L., Reiners, A., Christensen, U.R., Wicht, J.} 2013,
\href{http://adsabs.harvard.edu/abs/2013A\%26A...549L...5G}{\textit{A\&A}, 549, L5} 

\bibitem[Gastine \etal\ (2014)]{Gastine_etal14}
{Gastine, T., Yadav, R.K., Morin, J., Reiners, A., Wicht, J.} 2014,
\href{http://adsabs.harvard.edu/abs/2014MNRAS.438L..76G}{\textit{MNRAS}, 438, L76-80} 

\bibitem[Guerrero \etal\ (2016)]{Guerrero_etal16}
{Guerrero, G., Smolarkiewicz, P.K., de Gouveia Dal Pino, E.M., Kosovichev, A.G., \& Nansour, N.N.} 2016,
\href{http://adsabs.harvard.edu/abs/2016ApJ...819..104G}{\textit{ApJ}, 819, 104}

\bibitem[Howe (2009)]{Howe_09}
{Howe, R.} 2009,
\href{http://adsabs.harvard.edu/abs/2009LRSP....6....1H}{\textit{LRSP}, 6, 1}

\bibitem[Jeffers \etal\ (2018)]{Jeffers_etal18}
{Jeffers, S.V., Schoefer, P., Lamert, A., Reiners, A., Montes, D., Caballero, J.A., et al.} 2018,
\href{http://adsabs.harvard.edu/abs/2018arXiv180202102J}{\textit{accepted to A\&A, in press}, arXiv:1802.02102}

\bibitem[Jouve \etal\ (2010)]{Jouve_etal10}
{Jouve, L., Brown, B.P., \& Brun, A.S.} 2010,
\href{http://adsabs.harvard.edu/abs/2010A\%26A...509A..32J}{\textit{A\&A}, 509, A32}


\bibitem[Kapyla \etal\ (2013)]{Kapyla_etal13}
{K\"apyl\"a, P.J., Mantere, M.J., Cole, E., Warnecke, J., $\&$ Brandenburg, A.} 2013,
\href{http://adsabs.harvard.edu/abs/2013ApJ...778...41K}{\textit{ApJ}, 778, 41} 

\bibitem[Karak \etal\ (2017)]{Karak_etal17}
{Karak, B.B., \& Miesch, M.} 2017,
\href{http://adsabs.harvard.edu/abs/2017ApJ...847...69K}{\textit{ApJ}, 847, 69} 

\bibitem[Karak \etal\ (2015)]{Karak_etal15}
{Karak, B.B., K\"apyl\"a, P.J., K\"apyl\"a, M.J., Brandenburg, A., Olspert, N., \& Pelt, J.} 2015,
\href{http://adsabs.harvard.edu/abs/2015A\%26A...576A..26K}{\textit{A\&A}, 576, A26}

\bibitem[Metcalfe \etal\ (2017)]{Metcalfe_etal17}
{Metcalfe, T.S., \& van Saders, J.} 2017,
\href{http://adsabs.harvard.edu/abs/2017SoPh..292..126M}{\textit{Sol. Phys.}, 292, 126} 

\bibitem[Miesch \etal\ (2006)]{Miesch_etal06}
{Miesch, M.S., Brun, A.S., \& Toomre, J.} 2006,
\href{http://adsabs.harvard.edu/abs/2006ApJ...641..618M}{\textit{ApJ}, 641, 618}

\bibitem[Morris \etal\ (2017a)]{Morris_etal17}
{Morris, B.M., Hebb, L., Davenport, J.R.A., Rohn, G., \& Hawley, S.L.} 2017,
\href{http://adsabs.harvard.edu/abs/2017ApJ...846...99M}{\textit{ApJ}, 846, 99}


\bibitem[Nelson \etal\ (2014)]{Nelson_etal14}
{Nelson, N.J., Brown, B.P., Brun, A.S., Miesch, M.S., \& Toomre J.} 2014,
\href{http://adsabs.harvard.edu/abs/2014SoPh..289..441N}{\textit{Sol. Phys.}, 289, 441} 

\bibitem[Reiners \etal\ (2014)]{Reiners_etal14}
{Reiners, A., Sch\"ussler, M., \& Passegger, V.M.} 2014,
\href{http://adsabs.harvard.edu/abs/2014ApJ...794..144R}{\textit{ApJ}, 794, 144}

\bibitem[Reinhold \etal\ (2015)]{Reindhold_etal15}
{Reinhold, T., $\&$ Gizon, L.} 2015,
\href{http://adsabs.harvard.edu/abs/2015A\%26A...583A..65R}{\textit{A\&A}, 583, 65}

\bibitem[Schmidt \etal\ (2015)]{Schmidt_etal15}
{Schmidt, S.J., Hawley, S.L., West, A.A., Bochanski, J.J., Davenport, J.R.A., Ge, J., Schneider, D.P.} 2015,
\href{http://adsabs.harvard.edu/abs/2015AJ....149..158S}{\textit{AJ}, 149, 158}

\bibitem[Schuessler \etal\ (1996)]{Schuessler_etal96}
{Sch\"ussler, M., Caligari, P., Ferriz-Mas, A., Solanki, S.K., \& Stix, M.} 1996,
\href{http://adsabs.harvard.edu/abs/1996A\%26A...314..503S}{\textit{ApJ}, 314, 503}

\bibitem[See \etal\ (2016)]{See_etal16}
{See, V., Jardine, M., Vidotto, A.A., Donati, J.F., Boro Saikia, S.,Bouvier, J., et al.} 2016,
\href{http://adsabs.harvard.edu/abs/2016MNRAS.462.4442S}{\textit{MNRAS}, 462, 4442}

\bibitem[Shulyak \etal\ (2017)]{Shulyak_etal17}
{Shulyak, D., Reiners, A., Engeln, A., Malo, L., Yadav, R., Morin, J., \& Kochukhov, O.} 2017,
\href{http://adsabs.harvard.edu/abs/2017NatAs...1E.184S}{\textit{Natr. Astron.}, 1, 184} 

\bibitem[Strassmeier (2009)]{Strassmeier_09}
{Strassmeier, K.G.} 2009,
\href{http://adsabs.harvard.edu/abs/2009A\%26ARv..17..251S}{\textit{A\&A Rev.}, 17, 251}  

\bibitem[Strugarek \etal\ (2017)]{Strugarek_etal17}
{Strugarek,A., Beaudoin,P., Charbonneau, P., Brun, A.S., do Nascimento Jr., J.D.} 2017,
\href{http://adsabs.harvard.edu/abs/2017Sci...357..185S}{\textit{Science}, 357, 185} 

\bibitem[Viviani \etal\ (2017)]{Viviani_etal17}
{Viviani, M., Warnecke, J., K\"apyl\"a, M.J., K\"apyl\"a, P.J., Olpert, N., Cole-Kodikara, E.M., \etal\
} 2017,
\href{http://adsabs.harvard.edu/abs/2017arXiv171010222V}{\textit{submitted to A\&A}, arXiv:1710.10222}



\bibitem[Weber \& Browning (2016)]{Weber_etal16}
{Weber, M.A., \& Browning, M.K.} 2016,
\href{http://adsabs.harvard.edu/abs/2016ApJ...827...95W}{\textit{ApJ}, 827, 95} 

\bibitem[Weber \etal\ (2013a)]{Weber_etal13a}
{Weber, M.A., Fan, Y., Miesch, M.} 2013a,
\href{http://adsabs.harvard.edu/abs/2013ApJ...770..149W}{\textit{ApJ}, 770, 149}

\bibitem[Weber \etal\ (2013b)]{Weber_etal13b}
{Weber, M.A., Fan, Y., Miesch, M.} 2013b,
\href{http://adsabs.harvard.edu/abs/2013SoPh..287..239W}{\textit{Sol. Phys.}, 287, 239} 

\bibitem[Wright \etal\ (2013)]{Wright_etal13}
{Wright, N.J., Drake, J.J., Mamajek, E.E., Henry, G.W.} 2013,
\href{http://adsabs.harvard.edu/abs/2013AN....334..151W}{\textit{AN}, 334, 151}

\bibitem[Wright \etal\ (2016)]{Wright_etal16}
{Wright, N.J., \& Drake, J.J.} 2016,
\href{http://adsabs.harvard.edu/abs/2016Natur.535..526W}{\textit{Nature}, 535, 526}

\bibitem[Yadav \etal\ (2015)]{Yadav_etal15}
{Yadav, R.K., Gastine, T., Christensen, U.R., Reiners, A.} 2015,
\href{http://adsabs.harvard.edu/abs/2015A\%26A...573A..68Y}{\textit{A\&A}, 573, A68}

\bibitem[Yang \etal\ (2017)]{Yang_etal17}
{Yang, H., Jifeng, L., Gao, Q., Fang, X., Guo, J., Zhang, Y., et al.} 2017,
\href{http://adsabs.harvard.edu/abs/2017ApJ...849...36Y}{\textit{ApJ}, 849, 36}

\bibitem[Zhao \etal\ (2013)]{Zhao_etal13}
{Zhao, H., Bogart, R.S., Kosovichev, A.G., Duvall, T.L., \& Hartlep, T.} 2013,
\href{http://adsabs.harvard.edu/abs/2013ApJ...774L..29Z}{\textit{ApJL}, 774, L29}


\end{thebibliography}
\end{document}